\documentstyle[prl,aps]{revtex}
\input epsf.sty
\input psfig.sty
\newcommand{\vect}[1]{\mbox{\boldmath$#1$}}
\begin{document}
\draft
\twocolumn[\columnwidth\textwidth\csname@twocolumnfalse\endcsname
\title{Band Structures of $^{182}$Os Studied by GCM based on 3D-CHFB}
\author{Takatoshi Horibata$^{\rm a \, c}$
\footnote{e-mail: horibata@aomori.riken.go.jp},
Makito Oi$^{\rm b \, c}$, Naoki Onishi$^{\rm b \, c}$
and
Ahmad Ansari$^{\rm b \, d}$}
\address{$^{\rm a}$Department of Information System Engineering, 
Aomori University,Aomori-city, Aomori 030-0943, Japan}
\address{$^{\rm b}$Institute of Physics, College of Arts and Sciences, 
University of Tokyo, Komaba, Meguro-ku, Tokyo 153-8902, Japan}
\address{$^{\rm c}$Cyclotron Laboratory, Institute of Physical and 
Chemical Research(RIKEN), Hirosawa 2-1, Wako-city, Saitama 351-0198, Japan}
\address{$^{\rm d}$Institute of Physics, Doordarshan Marg, 
Bhubaneswar 751005, India}
\date{\today}
\maketitle
\begin{abstract}
Band structure properties of $^{182}$Os are investigated through a particle 
number and angular momentum constrained generator coordinate(GCM) calculation
based on self-consistent three-dimensional cranking solutions. 
From the analysis of the wave function of the lowest GCM solution, 
we confirm that this nucleus 
shows a tilted rotational motion in its yrast states, 
at least with the present set of force parameters 
of the pairing-plus-quadrupole interaction Hamiltonian. 
A close examination of behavior of other GCM solutions reveals a sign of a 
possible occurrence of multi-band crossing in the nucleus. Furthermore, in the 
course of calculations, 
we have also found a new potential curve along the prime 
meridian on the globe of the $J=18\hbar$ sphere. Along this new solution the 
characters of proton and neutron gap parameters get interchanged. Namely, 
$\Delta_p$ almost vanishes while $\Delta_n$ grows to a finite value close to the 
one corresponding to the principal axis rotation(PAR). 
A state in the new solution curve at the PAR point turns out 
to have almost the same characteristic features 
of an yrare $s$-band state which gets located just above the $g$-band in our 
calculation. This fact suggests a new type of seesaw vibrational mode of the 
proton and the neutron pairing, occurring through a wobbling motion. 
The mode is considered to bridge the $g$-band states and 
the $s$-band states in the backbending region. 


\end{abstract}
\addvspace{5mm}]

\narrowtext

\section{Introduction}
A limitation of a framework of the nuclear mean field approach whose 
validity is supported by the success of the idea of an independent 
particle motion has become one of the central matter of concern in 
recent years in nuclear physics. 
When we describe, for instance, high-$K$ states in a $\gamma$-soft nucleus 
or a large amplitude collective motion(LACM), such as a fission or
a process which leads to break up into fragmentation, which has recently 
attracted much attention through the molecular dynamics analysis, an 
assumption of stiffness against deformation is no longer valid. 
Furthermore, in the nuclear halo or in a skin structure only a small 
number of nucleons come into play in the region, and therefore the notion 
of a formation of a stable mean field will hardly be acceptable in such 
exotic situations.  For the description of these new topics one thinks of 
the two representative theories: the time dependent 
Hartree-Fock(TDHF) and the generator coordinate method\cite{HW53}.
They demonstrate their full ability in description of such situations.
Particularly, the GCM is fully quantum mechanical and it can be applied
to wide range of problems. Especially in the situations 
where several potential minima come into play in the motion. For instance,
LACM; in the situations of including a super deformation or a 
hyper deformation, it will be one of the promising methods because it 
expresses states as a superposition of many different product wave 
functions and can describe tunneling effects among the potential minima.
The basic idea of this method which goes beyond the conventional mean 
field theory, may well continue as one of the reliable frameworks in the 
coming century.

Since the nuclei in the region of osmium and tungsten generally exhibit
softness against the $\gamma$-deformation\cite{CHO88} 
they provide a nice testing ground for an approach which goes 
beyond the mean field theory.  We have been carrying out 
a systematic calculation based on a self-consistent three-dimensional 
cranked Hartree-Fock-Bogoliubov(3D-CHFB) theory in $^{182}$Os and 
have found a solution which suggests the existence of a tilted axis 
rotating(TAR) state\cite{HO94}. 
The resulting potential curve along the prime meridian on the globe of 
$J=18\hbar$ sphere has two TAR minima which locate symmetrically with respect 
to an equatorial plane.
Thus the GCM calculation by taking into account the north and the south
latitude as a generator coordinate to describe excited states is a highly 
interesting problem.

Our previous attempts based on the GCM calculations, however,
suffer from some difficulties\cite{HOO95}.  The first point was that in our 
calculation somewhat large energy splittings among the GCM solutions were 
obtained. Since we wish that our GCM solutions should correspond 
to the observed level spacings above the yrast states, the calculated energy 
spacings among these solutions should stay within the range of 10keV$\sim$100keV. 
The second difficulty in our calculation was that we hardly obtained any 
stable GCM solutions in the high angular momentum region. 

One of the problems in our approach is that we lose much information 
concerning the high frequency part of wave functions when we solve the GCM equation. 
Because in solving the GCM equation we have discarded the eigenstates of the norm 
overlap kernel whose eigenvalues are negligibly small. These states bring 
serious numerical errors when the equation is numerically solved\cite{RS80}. 
Therefore, they are considered to be unfavorable.
The other possible problem is that although the TAR potential spread 
non-locally over a surface of a globe with a constant angular momentum, we restricted the 
Hill-Wheeler integration to one-dimensional coordinate, that is we 
restricted the calculation only along the prime meridian. 
These drawbacks of our approach will be remedied by introducing the 
the angular momentum projection techniques\cite{OOTH98} 
and the two-dimensional Hill-Wheeler integration\cite{ACCREP1}.  
Since the programs for carrying out such calculations will become a project 
of inevitably exhaustive investigation, more basic study about the 
characteristic features of GCM solutions at this moment must be helpful. 

Generally, in the process of solving the GCM equation the expectation 
values of the angular momentum and the particle number may get shifted 
from the values of the CHFB constraints. For instance, the amplitudes of lower 
angular momentum components present in the cranking wave function will increase 
their probabilities in such a way to reduce the total energy of the system. 
To keep the correct configuration of wave functions we restrict the expectation 
values of the angular momentum and particle number to their given values during 
the process of solving the GCM equation. Applying the constrained GCM 
method\cite{Bo90} to the TAR potential curve to obtain the improved solutions 
is one of the main motivations of the present work.

By solving the constrained GCM equations 
in case of $^{182}$Os at $J=18\hbar$ we have found that the wave functions 
have symmetric or anti-symmetric 
properties with respect to the generator coordinate. This means that the 
broken symmetry of signature in TAR wave function has been recovered 
through the GCM procedure. Analyzing the wave function we have found that this 
nucleus must be a tilted-rotor in the present parameter set.  Furthermore, we 
have found the existence of the PAR components in the excited GCM solutions. 
This result is very interesting because we have solved the GCM equations based 
on the TAR potential which shows 
the minimum points located at a distance from the 
principal axis. This fact exhibits a sign of existence of a different band 
structure in the GCM solutions at $J=18\hbar$ state. We will discuss a 
possibility of multi-band crossing in this nucleus. Some preliminary results of 
our study are already presented in a short article\cite{HOO-I}. 

In the next section 2 a brief sketch of the formalism is presented. Results and 
discussions are presented in section 3. Finally section 4 contains a brief summary
and conclusions of the present investigations.

\section{Calculational method}

In this section we describe briefly the framework of our model. Details of the 
calculational methods for the 3D-CHFB and for the Hill-Wheeler equation are 
presented in our previous works\cite{HOO95,HO96,KO81}.

The model Hamiltonian used in the present study is the pairing plus $Q$-$Q$ force 
given by,
\begin{equation}
{\hat H} = \sum_{i} \epsilon_{i} c_{i}^{\dagger} c_{i}
 - \frac{1}{2} \kappa \sum_{\mu=-2}^{2}(-)^{\mu}{\hat Q}_{-\mu} {\hat Q}_{\mu}
 - \sum_{\tau=p, n} g_{\tau} {\hat P}_{\tau}^{\dagger} {\hat P}_{\tau}~,
\end{equation}
where ${\epsilon}_{i}$ represents the single particle energies and $c_{i}^{\dagger}$ 
and $c_{i}$ are the creation and the annihilation operators for particle states 
$\mid i \rangle$. Conditions $c_{i} \mid 0 \rangle = 0$ hold for all $i$, and 
$\mid 0 \rangle$ stands for the true vacuum. The parameters $\kappa$ and $g_{\tau}$ 
are the corresponding interaction strengths. The quadrupole operator is defined,
\begin{equation}
{\hat Q}_{\mu} = \sum_{ij} \langle i \mid r^2 Y_{2\mu}
 \mid j \rangle c_{i}^{\dagger} c_{j},
\end{equation}
as well as the pairing,
\begin{equation}
{\hat P}_{\tau}^{\dagger} = \sum_{i_{\tau} > 0}
  c_{i_{\tau}}^{\dagger}
  c_{\tilde{i_{\tau}}}^{\dagger}.
\end{equation}
This Hamiltonian is very simple for calculation but still effective in describing
essential features of the nuclear deformation and the pairing correlation.
 
In the present work the Hill-Wheeler equation is modified to include the constraint 
on the expectation values of the angular momentum and the particle number operators. 
As usual, first the norm overlap kernel is diagonalized,

\begin{equation}
  \int n(\psi , {\psi}^{\prime})\, u_{k}({\psi}^{\prime}) 
     {\rm d} {\psi}^{\prime} =
    {\nu}_{k}u_{k}(\psi),
\end{equation}
where $n(\psi ,{\psi}^{\prime})$ is Hermitian matrix defined as,

\begin{equation}
   n(\psi , {\psi}^{\prime})= \langle \Phi(\psi) \mid 
    \Phi({\psi}^{\prime}) \rangle.
\end{equation}
The wave function $\Phi(\psi)$ is the solution of the following 3D-CHFB equation
with eight constraints,

\begin{equation}
  \delta \langle \Phi \mid
     \left[
       {\hat H} - \sum_{k=1}^{3} \left( \mu_{k} {\hat J}_{k}
      + \xi_{k}{\hat B}_{k} \right) -  \sum_{\tau=p,n}
        \lambda_{\tau} {\hat N}_{\tau}
     \right] 
  \mid \Phi \rangle = 0 \, ,
\end{equation}
where $\mu_{k}$, $\xi_{k}$ and $\lambda_{\tau}$ are the Lagrange multipliers 
which are adjusted to satisfy the following constraints,
\begin{equation}
\langle \Phi \mid {\hat J}_{k} \mid \Phi \rangle = j_{k}
  \hspace{1.0cm} (k=1, 2, 3),
\end{equation}
\begin{equation}
  \langle \Phi \mid {\hat N}_{\tau} \mid \Phi \rangle = n_{\tau} \,
   \hspace{1.0cm} (\tau =p, n),
\end{equation}
and
\begin{equation}
  \langle \Phi \mid {\hat B}_{k} \mid \Phi \rangle = 0 \, 
   \hspace{1.0cm} (k=1, 2, 3),
\end{equation}
where ${\hat B}_{k}$'s express three off-diagonal elements of the 
mass-quadrupole tensor in Cartesian-coordinate representation,
\begin{equation}
  {\hat B}_{k}= \frac{1}{2}({\hat Q}_{ij}+{\hat Q}_{ji}) \, 
 \hspace{1.0cm}  (ijk; {\rm cyclic})\, .
\end{equation}
We know from our previous study that a calculation along the prime meridian always 
gives lower energy when compared with the calculation performed on the whole surface 
of the globe\cite{HO96}. 
Therefore, in the present study we restrict the generator coordinate only to the latitude 
$\psi$ along the prime meridian. Thereby the CHFB equations become now only two-dimensional.
The angle $\psi$ is measured with respect to the $x$-axis so that 
${\hat J}_1={\hat J} \cos \psi$, ${\hat J}_2=0$ and ${\hat J}_3={\hat J} \sin \psi$. 
This also helps us not to face any problem in choosing the phase in the square root which 
will appear in the calculation of eq.(5)\cite{HHR82}.
By utilizing the eigenfunctions $u_{k}({\psi)}$ of the overlap kernel, eq.(4),
we can construct the orthogonal sets which are called the natural 
states\cite{RS80}, 

\begin{equation}
   \mid k \, \rangle = \frac{1}{\sqrt{{\nu}_k}} \int u_{k}(\psi)
    \mid \Phi(\psi) \, \rangle {\rm d} \psi.  
\end{equation}

Generally, in the process of solving the GCM equation the expectation values 
of angular momentum and the particle numbers will somewhat get shifted from the values 
of the CHFB constraints. To keep the appropriate wave functions we restrict 
the expectation values of the scalar of squared angular momentum vector $\vect{J}^{2}$ 
and particle numbers to their correct values during the process of solving the GCM equations. 
Therefore we solve the following constrained Hill-Wheeler equation,

\begin{equation}
\sum_{k^{\prime}} \left[ H_{k,k^{\prime}}-\lambda_{\tau}^{(\alpha)}N^{\tau}_{k,k^{\prime}}
             -{\mu}^{(\alpha)} \vect{J}^{2}_{k,k^{\prime}}\right]
        g_{k^{\prime}}^{(\alpha)}  =     
       E^{(\alpha)}g_{k}^{(\alpha)}.  
\end{equation}
The Lagrange multipliers $\lambda_{\tau}^{(\alpha)}$ and ${\mu}^{(\alpha)}$ in eq.(12)
control the expectation values of particle numbers and the square of angular momentum 
in the GCM wave functions and are different from those appearing in eq.(6). 
The superscript ``$\alpha$'' indicates the index of the eigenstates. 
The matrices $H_{k,k^{\prime}}$, $N^{\tau}_{k,k^{\prime}}$ and $\vect{J}^{2}_{k,k^{\prime}}$ 
are given as follows,

\begin{equation}
     H_{k,k^{\prime}} =
 \int {\rm d} \psi {\rm d} {\psi}^{\prime} \frac{u^{\ast}_k(\psi)}
                                                                {\sqrt{{\nu}_k}}
            \langle \Phi(\psi) \mid {\hat H} \mid \Phi({\psi}^{\prime}) \rangle
       \frac{u_{k^{\prime}}({\psi}^{\prime})}{\sqrt{{\nu}_{k^{\prime}}}},
\end{equation}

\begin{equation}
     N^{\tau}_{k,k^{\prime}} =
 \int {\rm d} \psi {\rm d} {\psi}^{\prime} \frac{u^{\ast}_k(\psi)}
                                                                {\sqrt{{\nu}_k}}
        \langle \Phi(\psi) \mid {\hat N}_{\tau} \mid \Phi({\psi}^{\prime}) \rangle
       \frac{u_{k^{\prime}}({\psi}^{\prime})}{\sqrt{{\nu}_{k^{\prime}}}},
\end{equation}
and 
\begin{equation}
     \vect{J}^{2}_{k,k^{\prime}} =
 \int {\rm d} \psi {\rm d} {\psi}^{\prime} \frac{u^{\ast}_k(\psi)}
                                                                {\sqrt{{\nu}_k}}
        \langle \Phi(\psi) \mid {\hat {\vect{J}^{2}}} \mid \Phi({\psi}^{\prime}) \rangle
       \frac{u_{k^{\prime}}({\psi}^{\prime})}{\sqrt{{\nu}_{k^{\prime}}}}.
\end{equation}
Detailed expressions for the matrix elements appearing in the integration
are given in Ref. \cite{HOO95}. 
The eigenstates of our GCM equation are given by,

\begin{equation}
    \mid \Psi_{\alpha} \, \rangle = \sum_{k,{\nu}_{k} \neq 0} g_{k}^{(\alpha)} \mid k \rangle,
\end{equation}
where $g_k^{(\alpha)}$ are obtained as solutions of eq.(12).
From eq.(11) we find the corresponding weight function,
\begin{equation}
    f^{(\alpha)}(\psi) = \sum_{k,{\nu}_{k} \neq 0} g^{(\alpha)}_{k} u_{k}(\psi).
\end{equation}
Then, finally the physical quantities are calculated using the matrix elements in eq.(13-15)
and the wave function $g_k^{(\alpha)}$ as follows,

\begin{equation}
 \left( \begin{array}{c}
     H^{(\alpha)}        \\ \\
     N_{\tau}^{(\alpha)} \\ \\
     J^{2(\alpha)}
        \end{array} \right) =
   \sum_{k,k^{\prime}} g^{\ast (\alpha)}_k 
    \left( \begin{array}{c} 
        H_{k,k^{\prime}}  \\ \\ 
        N^{\tau}_{k,k^{\prime}} \\ \\ 
        \vect{J}^{2}_{k,k^{\prime}}
           \end{array} \right)
              g^{(\alpha)}_{k^{\prime}}.
\end{equation}

\section{Numerical results and discussions}

\subsection{Solution of 3D-CHFB equations}

Recently a systematic calculation based on the 3D-CHFB theory revealed the 
possible existence of a tilted axis rotating potential in $^{182}$Os
nucleus\cite{HO94,HO96}. In our previous calculations, however, 
the force parameters were chosen so as to reproduce just the overall features 
of low-angular momentum states of nuclei in this mass region. The observed first 
backbending in A$\sim$180 region is expected to be due to $i_{13/2}$ 
neutron alignments, whereas in our earlier calculation protons were playing the 
role. The reason was attributed to our slightly too strong Q$\cdot$Q 
force parameter and also to rather strong pairing interaction for neutrons. 
In the present calculation  we reduced the Q$\cdot$Q 
force parameter and the strength of neutron pairing interaction so as to
make rotation alignment of neutron occur in the PAR states, 
which is considered to be yrast in $^{182}$Os. Our new force parameters
are chosen so as to give as $\beta=0.280$, $\Delta_p=1.040$MeV and $\Delta_n=0.925$MeV
in the ground state. 
Moreover, we have attained considerable high precisional calculation this time.
We have increased the number of iteration to achieve better convergence when solving 
by the steepest descent method. 
We have fixed an increment of angular momentum as $\Delta J=0.1\hbar$ in the 
calculation along the yrast line up to $J=30\hbar$ and set the maximum iteration 
for convergence as 300 times and the minimum iteration as 50 times for each step. 
In our previous calculation, those iteration times were taken as 120 and 30, respectively. 
The size of the model space and the single particle energies $\epsilon_{i}$ in the
Hamiltonian are not changed from our previous study. 
They are shown in Table 1 together with quantum numbers such as isospin, radial 
node number, orbital angular momentum and total angular momentum\cite{GL67}.

In a similar manner as the previous papers, we obtained PAR solutions
by cranking up along $x$-axis perpendicular to the symmetry axis
of the non-rotating state and also by cranking down along the axis.
The two bands cross at $J\sim20\hbar$ in the present case. At the crossing point in
the $g$-band the size of proton gap parameter is $\Delta_p=0.87$MeV and the 
neutron gap parameter almost vanishes. The characteristic features of the yrast line
are indicated in Fig.1. 
\vspace{3mm}

 \begin{tabular}{lc}
   \multicolumn{2}{c}{\bf Model space and single particle energies} \\
        \\
   \multicolumn{2}{l}{proton} \\ \hline
   \multicolumn{1}{l}{orbits} & \multicolumn{1}{c}{energy (MeV)} \\ \hline
        $1f_{7/2}$   &  \ 8.06435 \\
        $0i_{13/2}$  &  \ 7.55675 \\
        $0h_{9/2}$   &  \ 7.23480 \\
        $2s_{1/2}$   &  \ 3.87120 \\
        $1d_{3/2}$   &  \ 3.59468 \\
        $0h_{11/2}$  &  \ 2.16538 \\
        $1d_{5/2}$   &  \ 1.29040 \\
        $0g_{7/2}$   &  \ 0.64520 \\
        $0g_{9/2}$   &   -3.50251 \\ \hline
     \\
   \multicolumn{2}{l}{neutron}    \\ \hline
   \multicolumn{1}{l}{orbits} & \multicolumn{1}{c}{energy (MeV)} \\ \hline
        $0i_{11/2}$  &  \ 7.55740 \\
        $1g_{9/2}$   &  \ 6.74629 \\
        $2p_{1/2}$   &  \ 4.40579 \\
        $1f_{5/2}$   &  \ 3.39191 \\
        $2p_{3/2}$   &  \ 3.02322 \\
        $0i_{13/2}$  &  \ 1.56626 \\
        $0h_{9/2}$   &  \ 0.82954 \\
        $1f_{7/2}$   &  \ 0.16591 \\
        $0h_{11/2}$  &   -4.23988 \\ \hline
 \end{tabular}

\vspace{0.2cm}
\noindent
Table 1: Model space and the single particle energies in the present calculation. 
In the space 36 valence particles and 40 core particles are taken for the proton and also 
36 valence particles and 70 core particles are considered for the neutron.  

\vspace{0.7cm}


\begin{figure}
\psfig{figure=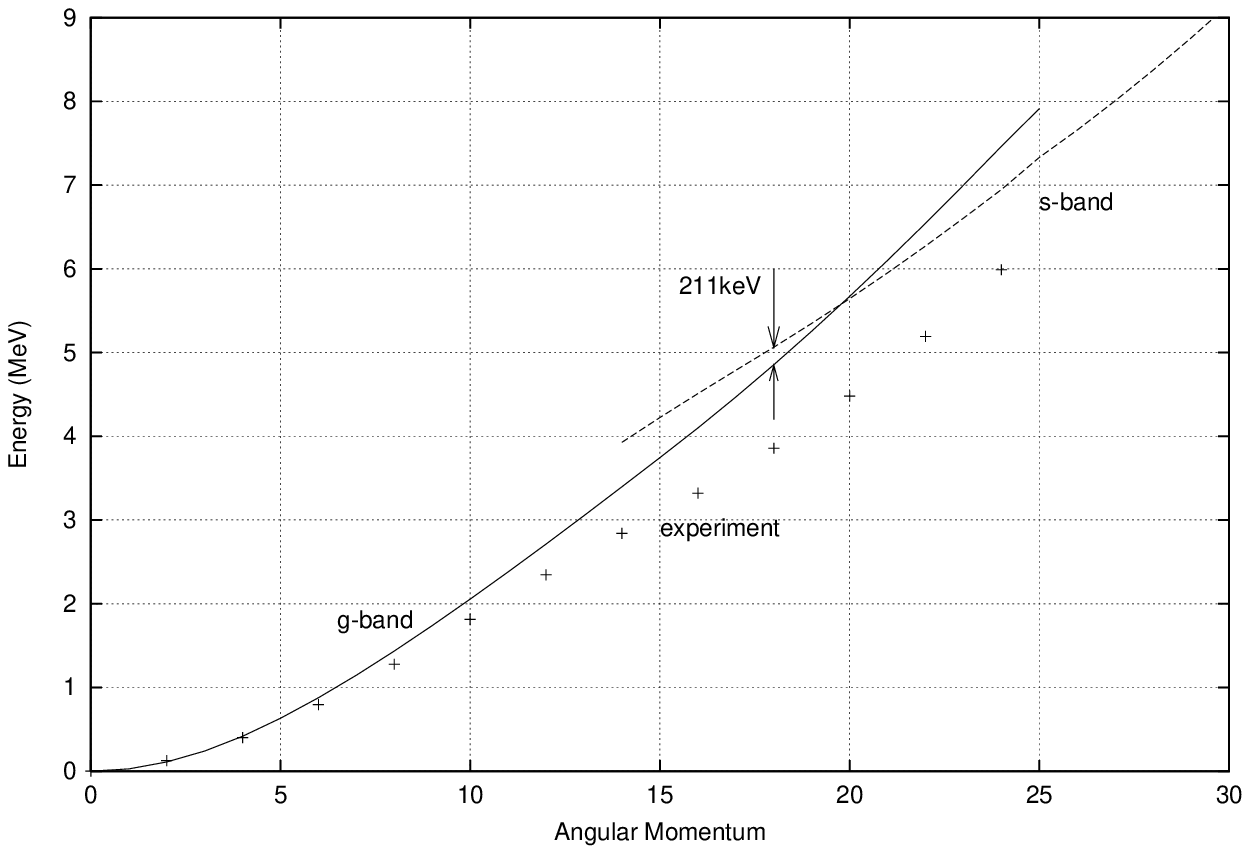,width=8.5cm}
\psfig{figure=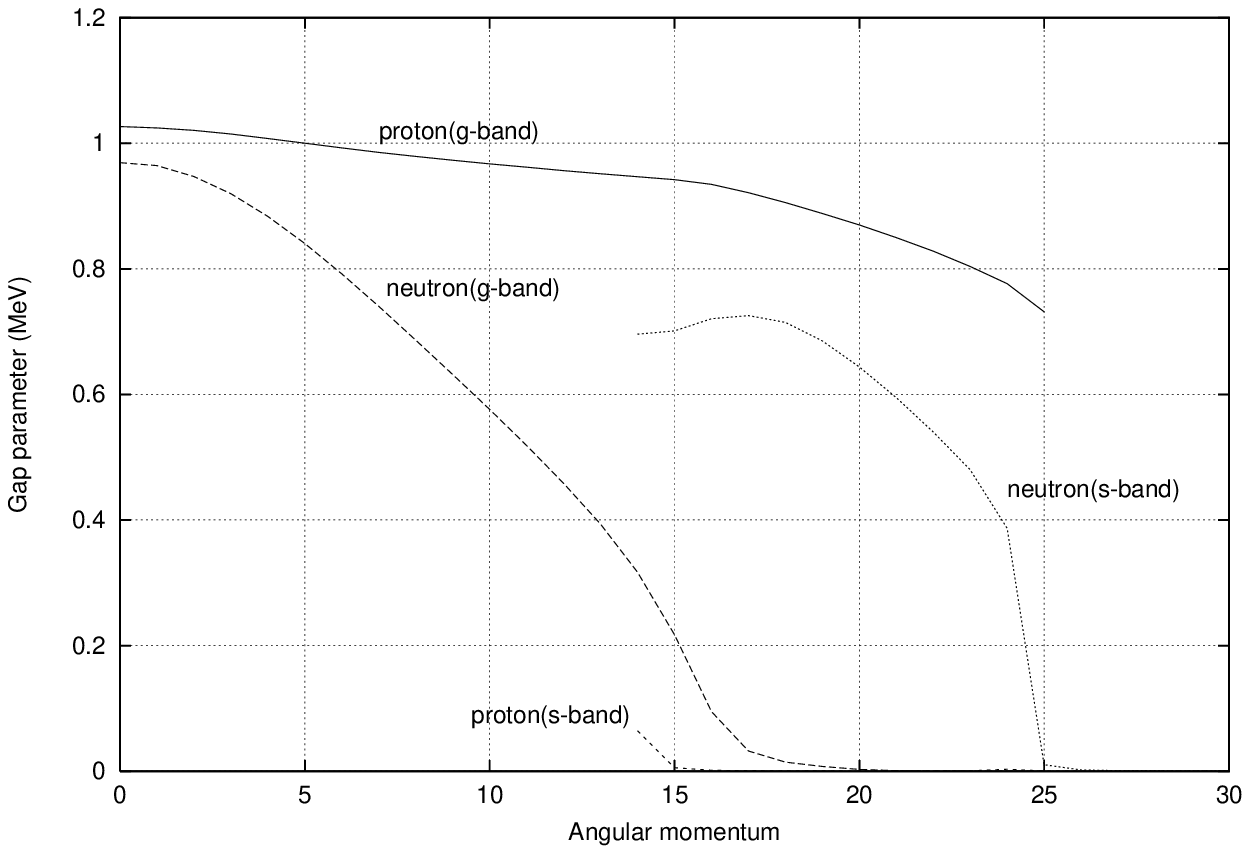,width=8.5cm}
\psfig{figure=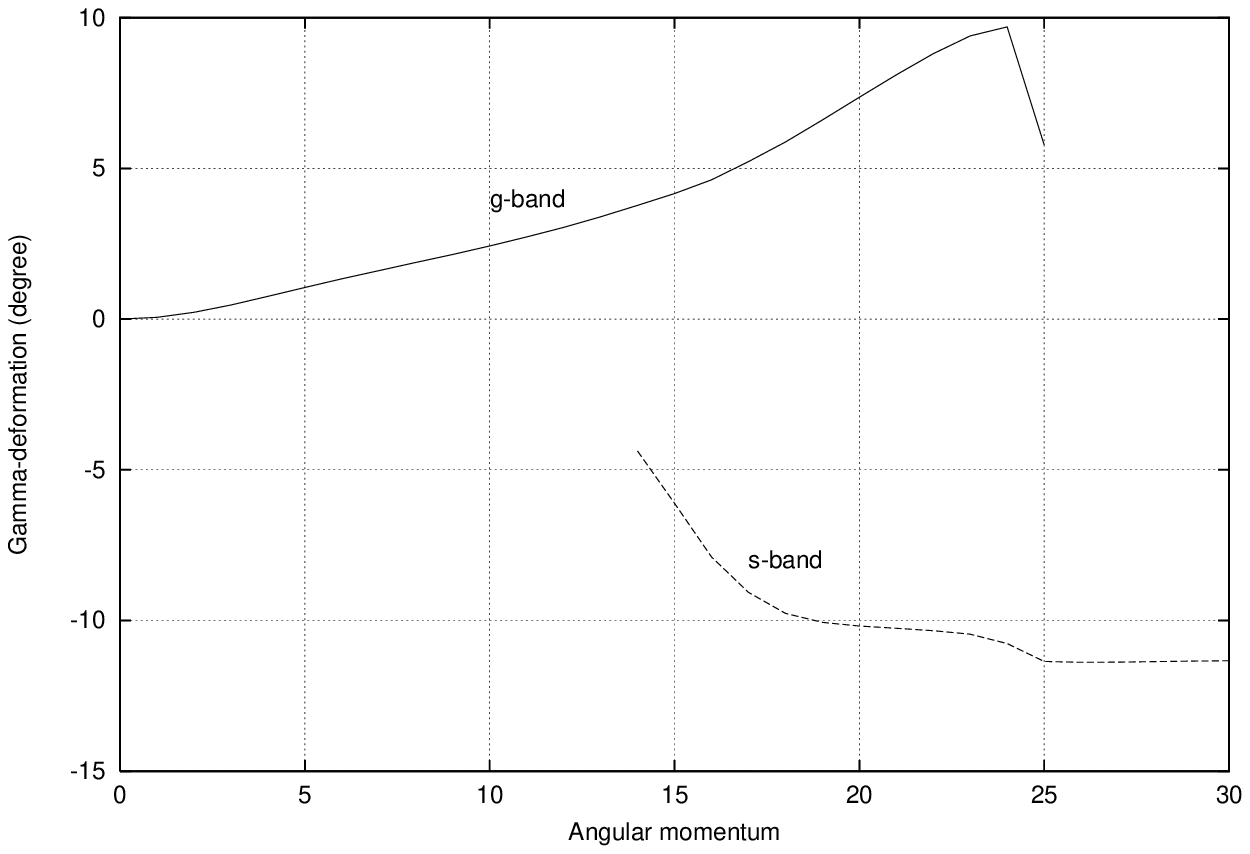,width=8.5cm}
\caption{Fig.1a shows the behavior of energy curves of the $g$-band and the 
$s$-band as well as the experimental value. 
The two bands cross at $J\sim20\hbar$ in our calculation. 
The energy value 211keV indicated in the figure shows the energy 
difference between the $J=18\hbar$ yrare state and the yrast state. 
Fig.1b shows the behavior of the gap parameters for each band.  
The change of $\gamma$-deformation is indicated in Fig.1c.}
\end{figure}

Fig.1a shows the behavior of energy curves of the $g$-band and the $s$-band as well as 
the experimental value taken from Ref.\cite{CHO88}.  In our previous 
calculation\cite{HO96} both bands cross at $J=14\hbar$ due to the proton
alignment. In the present case, after the band crossing at $J=20\hbar$ the 
$g$-band continues up to $J=25\hbar$ state. We can not obtain any $g$-band components 
above this state and after the intensive iteration processes the CHFB solution finally 
converges to a next state which belongs to the $s$-band.
In the same manner, the $s$-band obtained by cranking down calculation continues down to
$J=14\hbar$ state. When decreasing the angular momentum by a unit $0.1\hbar$ from this state
the solution converges to a new state in the $g$-band after many iteration
processes. In Fig.1b we show the behavior of the gap parameters. The strength of the 
neutron gap parameter both in the $g$-band and the $s$-band are much reduced when they are
compared with our previous calculation\cite{HO96}. In the present calculation, 
the proton gap parameter almost keeps its value until the end of the $g$-band.  
The change of $\gamma$-deformation is indicated in Fig.1c. The $\gamma$-deformation changes
drastically at the band crossing point. The value in the $g$-band is $\gamma_g = 7.36^{\circ}$
and in the $s$-band is $\gamma_s = -10.19^{\circ}$ at the point.

We tilted rotating axis along the prime meridian, which intersects with $z$- and $x$-axis 
on the sphere of $J=18 \hbar$. As in our previous calculation, here too the existence of 
a stable TAR state is confirmed in the region of $\psi = 20^{\circ} \sim 30^{\circ}$ as 
shown in Fig.2a.


The TAR minimum takes place at $\psi=24^{\circ}$, and its depth is 
$V_{TAR}=-0.166$MeV. 
We can find an occurrence of phase transition between the PAR and the TAR point at about
$\psi=10^{\circ}$. Starting from the PAR the potential curve ends at this point 
with its energy $V_{first}=0.189$MeV. 
After this point, a new potential curve with different configuration appears which 
continues up to $\psi=42^{\circ}$.
In the region between $\psi=10^{\circ}$ and $ \psi=42^{\circ}$, in which the TAR appears,
the proton gap parameter almost vanishes and the neutron gap parameter recovers its
original strength of the ground state value as shown in Fig.2b. This fact implies that 
the TAR is generated due to a collapse of the proton gap parameter in our calculation.  
Thus the  phase transition
is considered to be caused by a crossing of a band with $\Delta_n=0$ and $\Delta_p \ne 0$ 
to an another band with $\Delta_n \ne 0$ and $\Delta_p=0$.

We encounter the next phase transition point at $43^{\circ}$ with $V_{second}=0.470$MeV. 
At this point the configuration of wave function also changes drastically. That is, the
proton gap parameter recovers its strength and oppositely the neutron gap parameter
reduces its value to about 70$\%$ of the value at the TAR region as shown in Fig.2b.
In the region between $\psi=44^{\circ}$ and $\psi=52^{\circ}$ both the proton and
the neutron gap parameters take about $\Delta_p \sim \Delta_n \sim 0.5$MeV.
Calculation above this region, where we did not explore in our previous GCM study, 
the potential curve shows relatively smooth behavior and it goes down into negative 
value at the north pole.  The energy value at the point is $V_{n-pole}=-0.167$MeV. 
This means that the states around the north pole are lower in energy compared with 
the PAR state. The wave function of this state consists of the positive parity $i_{13/2}$ 
neutron which align their angular momenta to the symmetry axis and contribute 
to $K=8$ band state.
At this moment, we do not have any experimental evidences for the TAR minimum or the 
existence of energetically lower states near the north pole.  
\noindent
\begin{figure}
\psfig{figure=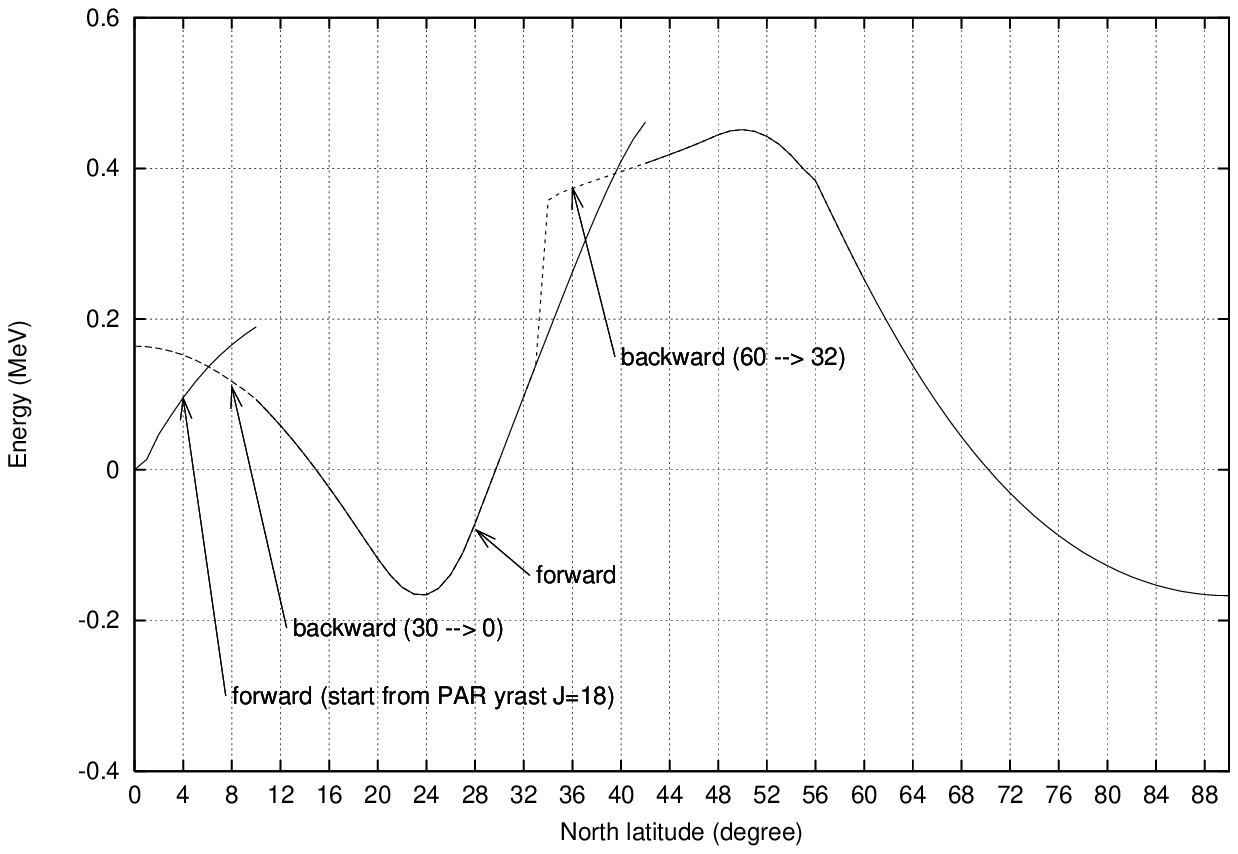,width=8.5cm}
\psfig{figure=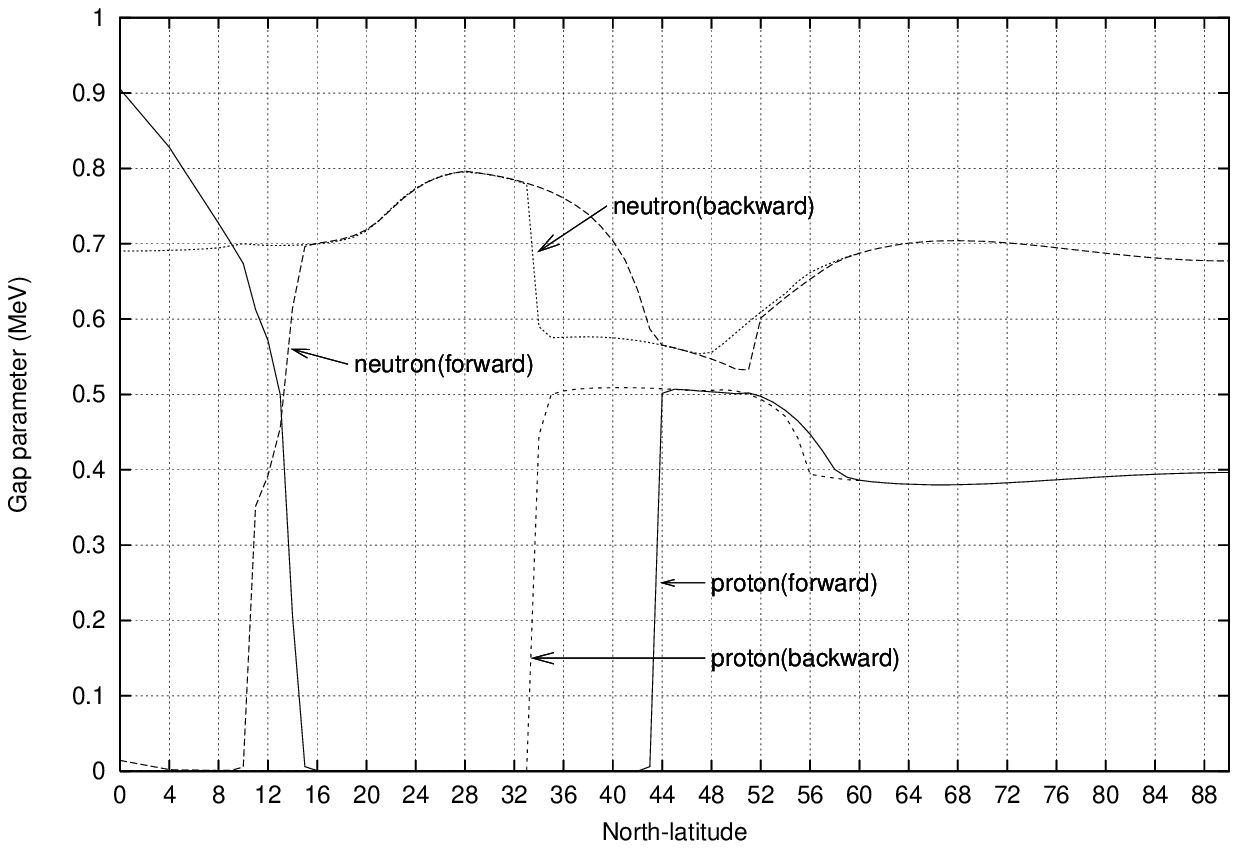,width=8.5cm}
\caption{Fig.2a shows the energy curves along the prime meridian on the globe 
of $J=18\hbar$ sphere of $^{182}$Os. The solid curves (indicated ``forward'') are obtained 
by increasing the north latitude angle starting from $\psi=0^{\circ}$ up to 
$\psi=90^{\circ}$. The two phase transitions are observed at $\psi=10^{\circ}$ 
and $\psi=42^{\circ}$. The two dashed curves show the another 3D-CHFB solutions (indicated 
``backward'') which are obtained by decreasing the angle. Along this new solutions the 
characters of the proton and the neutron gap parameters mutually get interchanged as seen 
in Fig.2b. All the energy curves are mirror symmetric with respect to an equatorial plane 
of the globe.
Fig.2b shows behavior of the pairing gap parameters for the proton and the neutron along 
the curves indicated in Fig.2a. We can see that the TAR potential is caused by a sudden 
decrease of proton gap parameter and the simultaneous sudden restoration of the neutron 
gap parameter. Near the PAR point the proton gap parameter $\Delta_p$ almost vanishes and 
the neutron gap parameter $\Delta_n$ recovers its value.}
\end{figure}
\noindent
The GCM study based on this full potential curve will certainly give some 
clues to understand the new dynamics characteristic in this mass region. 
Unfortunately, our previous calculation were restricted to a region up to the second 
phase transition point\cite{HOO95}.

On the other hand, we show in Fig.2a the existence of another 3D-CHFB 
solutions along the prime meridian in this nucleus\cite{HO97}. 
The first solutions have been obtained by a calculation starting from the point 
$\psi=30^{\circ}$ back to the point $\psi=0^{\circ}$. The next solutions 
starting from the point $\psi=62^{\circ}$ back to the point $\psi=32^{\circ}$. 
Along this new first solution the characters of proton and neutron gap parameters 
mutually get interchanged as shown in Fig.2b. Namely, $\Delta_p$ almost 
vanishes while $\Delta_n$ grows up to a finite value at the PAR point. Therefore 
this new solution has an $s_p$-band character. At the PAR point, the excitation
energy and the gap parameters $\Delta_p$ and $\Delta_n$ of this new solution takes 
the close value as that of the yrare $s$-band state at $J=18\hbar$. The excitation 
energy of the new solution at the PAR point is $E_{tilt-back} \sim$ 164keV whereas the 
energy difference between an yrast $g$-band state and an yrare $s$-band state at 
$J=18\hbar$ is $E_{g-s} \sim$ 211keV in our calculation(see Fig.1a). The energy 
difference between the states in the excitation energy is about 47keV. 
At this moment, we think that this difference 
arises due to still insufficient convergence during our iteration process. 
If we start a tilting calculation along the prime meridian from the $E=211$keV 
$J=18\hbar$ yrare state, a resulting potential curve just after the first phase transition 
point at $\psi = 10^{\circ}$(see Fig.2a) appears very close to the original potential 
curve which is obtained by starting from the yrast $J=18\hbar$ PAR state. 
The new state at the PAR point can naturally be 
considered as a member of an yrare $s$-band states, but more precise study will
be necessary about this point. 
The value of gap parameters at the PAR point in the new solution are almost the same 
as that of the corresponding yrare state in the $s$-band.

In the present calculation the $s$-band solutions in the PAR calculation continue down 
to a state with angular momentum $J=14\hbar$. Along the solution curve starting inversely 
from the $J=30\hbar$  the neutron gap parameter gradually increase up to
$\Delta_n=0.76$MeV at $J=16\hbar$.  
The $\gamma$-deformation changes drastically between the yrast and the yrare states 
at $J=18\hbar$. The value of the yrast state is $\gamma=5.88^{\circ}$ and that of the 
yrare state is $\gamma=-9.77^{\circ}$. From these results, we can say that a new type 
of seesaw vibrational mode of the proton and the neutron pairing, occurring through 
wobbling motion accompanying the large $\gamma$-oscillation, can bridge the yrast 
state in the $g$-band and the yrare state in the $s$-band in the backbending region of
$^{182}$Os.

\subsection{Solving GCM equations}

We have solved the constrained Hill-Wheeler equation eq.(12) for $^{182}$Os 
at $J=18\hbar$ in which the north and the south latitude angles are taken 
as the generator coordinate. The integration has been carried out in the 
range of $\psi=-90^{\circ}\sim90^{\circ}$. In numerical calculations 
we replace the integration over the range into a discrete summation with a step of 
$4^{\circ}$ starting from  $\psi=-88^{\circ}$ to $88^{\circ}$ in which totally 45 
integration points are contained. The size of the step should be chosen carefully. 
In the GCM calculation, we first diagonalize the overlap kernel to construct a 
complete orthogonal set of the GCM wave function.
If we set the size to smaller than $4^{\circ}$ in the calculation, some eigenvalues 
of the norm overlap kernel turn negative because of an over-completeness of the 
overlap kernel. In our calculation all eigenvalues of the norm kernel stay 
positive and the smallest one is in the order of $\sim10^{-7}$ which stays in
the range of numerical error.

In this study we have carried out two types of calculations. The first type of calculation 
is the numerical solutions without any constraints whose solutions are shown in Fig.3a. 
The second one is the solutions in which the expectation values of the squared angular 
momentum and the particle number operators are constrained. The results are shown in 
Fig.3b. The lowest ten excited states are shown in both the figures.
The abscissa of the figures are the number of eigenstates taken into the calculation 
of eq.(12) which are  obtained from a diagonalization of the norm overlap kernel.
The eigenstates are taken  in the decreasing order of the size of eigenvalues of eq.(4).


 In Fig.3a, we can see that the solutions do not show any stable plateau 
in the whole range of abscissa coordinates. The situation is not improved at all
when compared to our previous works.  We only find relatively small plateau in the 
region of lower excitation energies within a range of abscissa coordinate $35\sim40$. 
The lowest two solutions in the region are almost degenerate and the next higher 
state lies more than 700keV higher in the energy position.  In the previous 
calculation we also obtained relatively large energy splittings among the GCM solutions
and we attributed the reason 
to relatively deep potential of the TAR. The value was $V_{TAR}^{old}=-0.367$MeV 
whereas in the present case the value is $V_{TAR}=-0.166$MeV. 
The depth is much smaller this time and the value is less than half of the previous 
one. From this fact, we can consider that only the shallowness of the TAR 
potential may not be enough to suppress the size of energy splittings among the states.

In Fig.3b, we can find an existence of relatively large and stable plateaus in the whole 
range of the abscissa coordinate. 
The energy splittings among the states are in the range of 10keV $\sim$ 100keV. 
These values are reasonable for an interpretation of any excited states observed 
in this mass region. At least the lowest seven states consists relatively
stable plateau within the range of abscissa coordinate $30\sim35$(here after we refer this
region as {\em regular}). From the present investigation, we understand that the 
conservation of the square of angular momentum and the particle number 
can very much increases the stability of the GCM solutions. 
Usually a large angular momentum fluctuation is included in the wave function 
resulting from a cranking model. Since this value is $\Delta J^2 \sim 100{\hbar}^2$ 
in our case, we restrict the expectation value of $J^2$ to $\sim 420{\hbar}^2$ 
for studying the $J=18\hbar$ state.
In solving the Hill-Wheeler equation the expectation value of the total angular 
momentum does not remain constant when increasing the number of states but drifts to lower 
angular momentum value to reduce the total energy of the system. This situation can be 
clearly seen in Fig.4a. 
\noindent
\begin{figure}
\psfig{figure=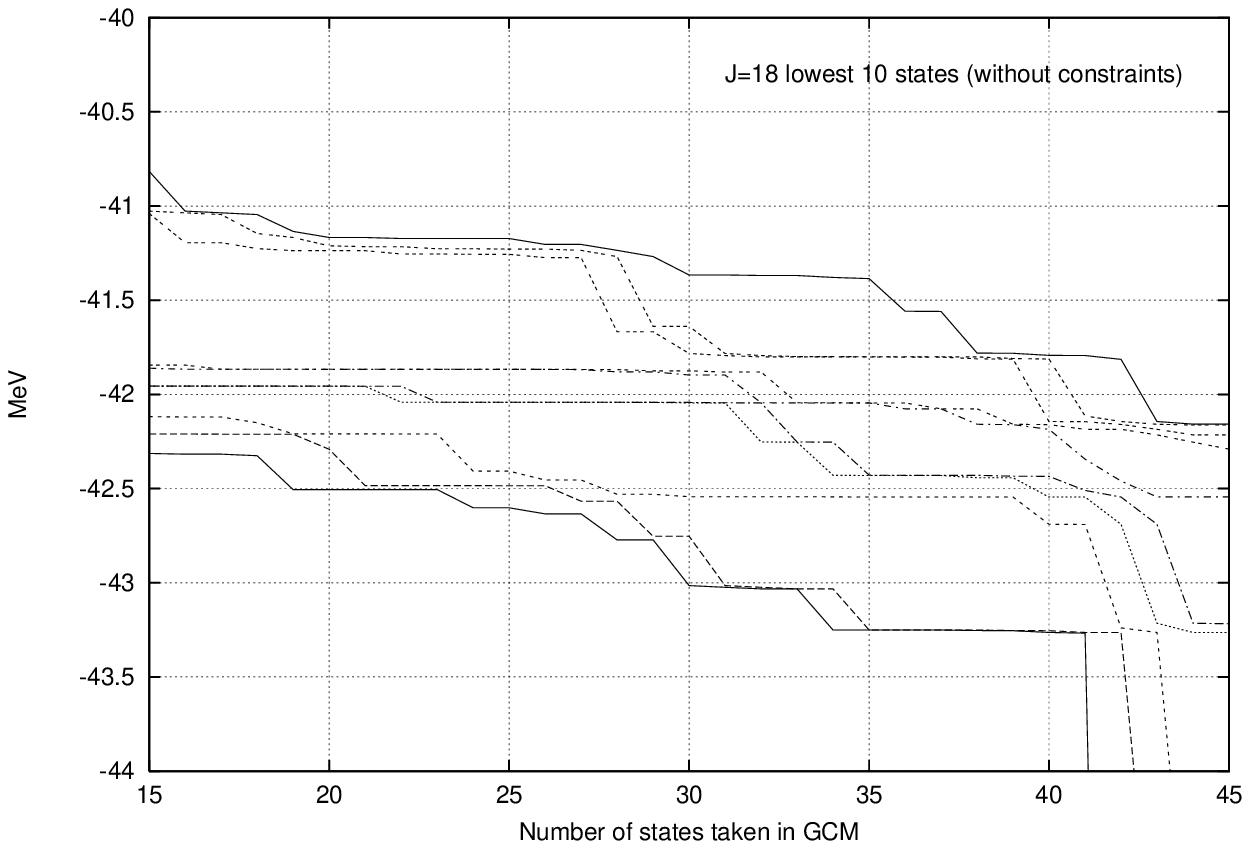,width=8cm}
\psfig{figure=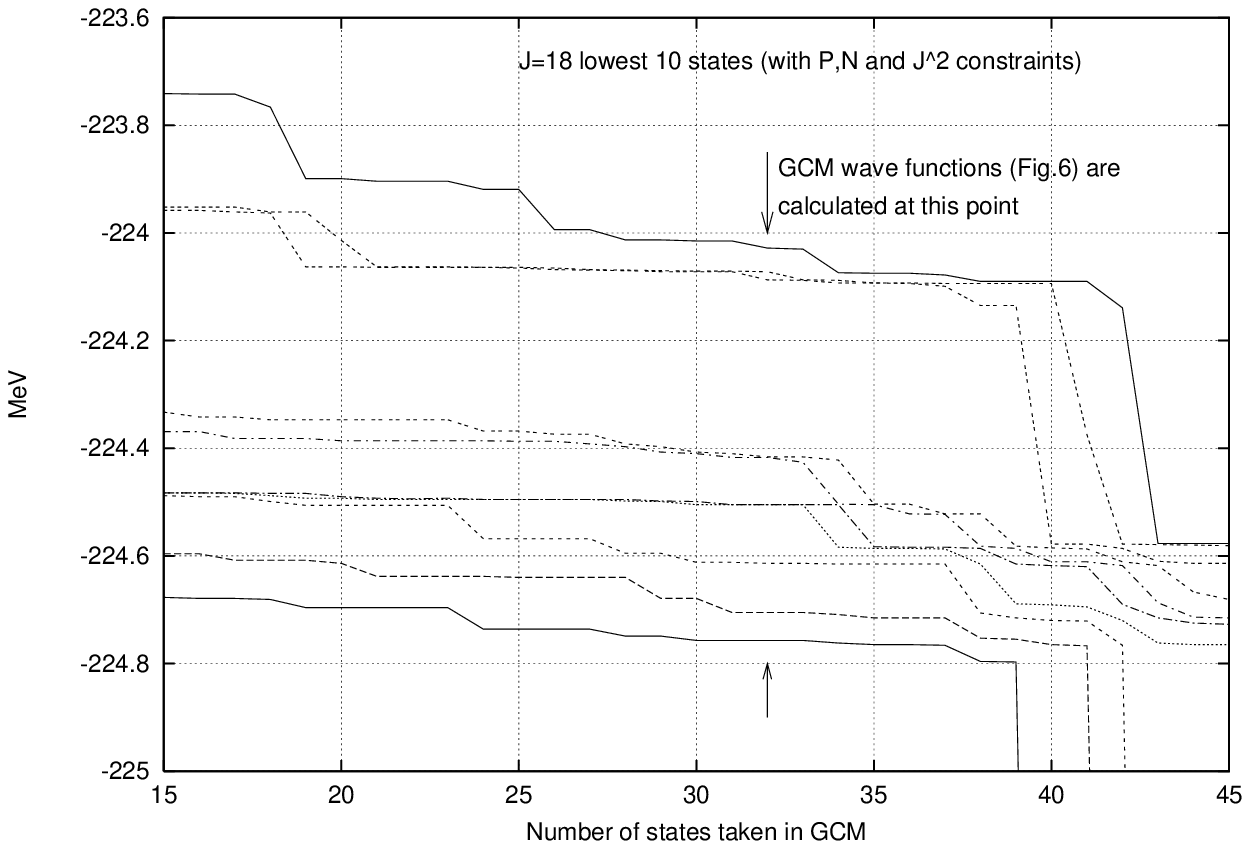,width=8cm}
\caption{Several lowest solutions of Hill-Wheeler equations. In the figures the 
abscissa indicates the number of solutions of eq.(4) considered in solving the equations. 
Fig.3a shows the solutions without any constraints and Fig.3b indicates the results with 
particle numbers and the square of angular momentum are constrained(see eq.(12)). Since the
wave function in the cranking model suffers from a large angular momentum fluctuation, we 
took the constrained value of the squared angular momentum value as 
$J^2 \sim 420 \hbar^2$
in the calculation for the $J=18\hbar$ state. We can see from the figure that 
the constraints very much contribute to stabilize the GCM solutions. In Fig.3b we indicate the
point at which GCM weight functions eq.(17) are calculated(see Fig.6).}
\end{figure}
\noindent
The solutions without any constrains do not keep the expectation
value of angular momentum. Contamination of lower angular momentum components into our GCM
solutions may force the excitation energy of each solution to lower values. This is the
main reason for obtaining the unstable GCM solution when solving with no constraints(Fig.3a). 
The angular momentum constraint will not be necessary if good angular momentum states
are used for the GCM calculation.


\begin{figure}
\psfig{figure=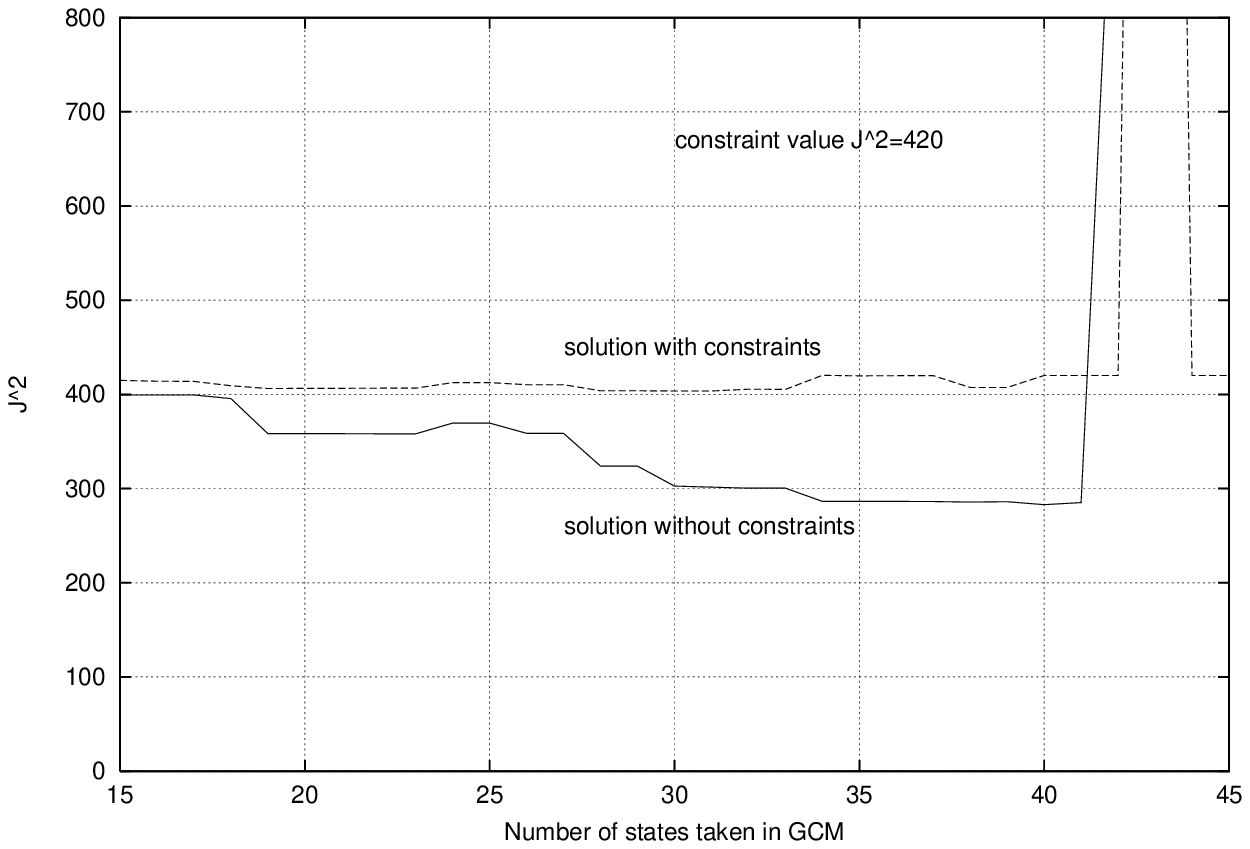,width=8cm}
\psfig{figure=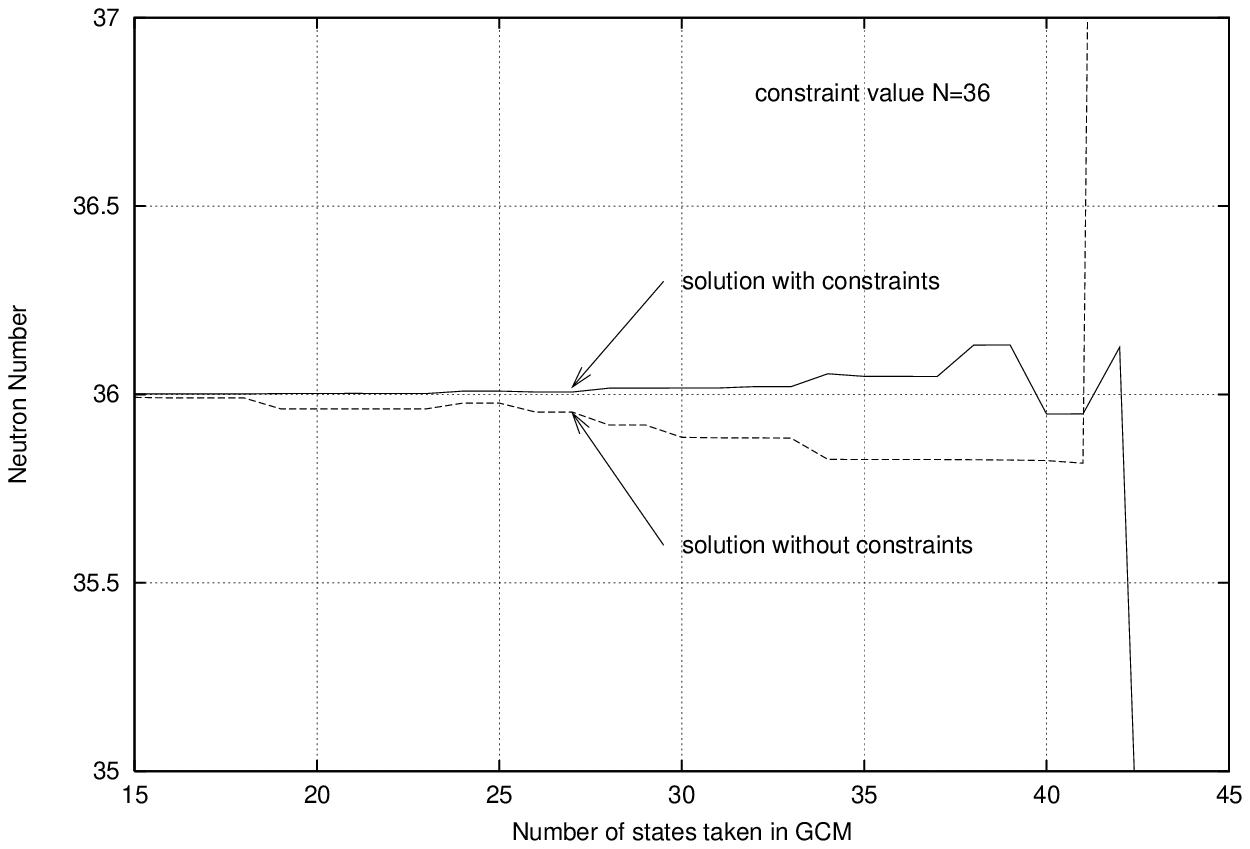,width=8cm}
\psfig{figure=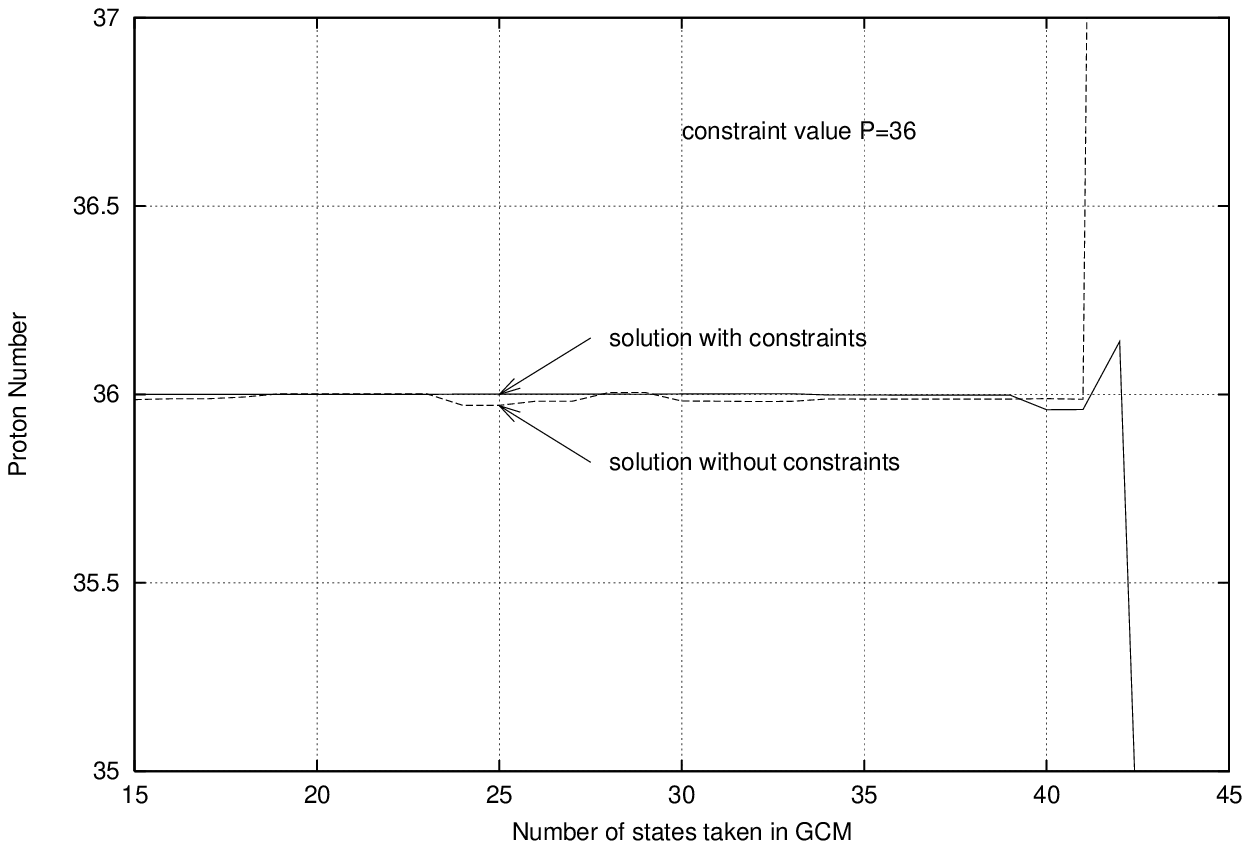,width=8cm}
\caption{Comparison of behaviors of the constrained values when solving the 
Hill-Wheeler equation with and without the constrained terms. Fig.4a indicates the difference 
of $J^2$ between the two results. We can see that the decrease of 
the expectation value of 
angular momentum in Fig.4a corresponds to the instability of the Hill-Wheeler equation shown 
in Fig.3a. Fluctuation for the neutron numbers(Fig.4b) and for the proton numbers(Fig.4c) are 
small in the present study.}
\end{figure}

In our present study constraints for the particle numbers do not appear to be essential.
Following the above discussions, contamination of more particle number components in the
wave function may lower the total energy of the system. But this tendency may be
suppressed by the contamination of lower angular momentum component in the wave function.
In our calculation, the fluctuations of particle numbers in the solutions of GCM equations 
without the constraints are negligibly small, that is, the fluctuation for the neutron is 
less than 0.56\%(Fig.4b) and for the proton the value is almost unchanged(Fig.4c).

Fig.5 shows the probability distribution of the GCM solutions. In the plane the radius
of each circle indicates the size of eigenvalue of the equation of norm overlap kernel, eq.(4), 
which express the probability of each solution. These solutions are taken into account in 
solving the GCM equations. For instance, the slope of a line for an energy versus $J^2$ 
plot(Fig.5a) indicates a value for the chemical potential $\mu^{(1)}$ in the eq.(12). 
In our calculation
this value becomes $6.20\times10^{-3}$ within the {\em regular} region. From the figure
we can see that the solutions are distributed approximately along this line.  
The most probable GCM states localize around the region $J^2=400{\hbar}^2 \sim 450{\hbar}^2$ 
and their energies are about $E^{\prime}\sim -42$MeV.

Fig.5b and Fig.5c show the distribution for the GCM solutions in the energy versus particle 
number plane. We can see the most probable states locate around the 
value of 36 both for the protons as well as for the neutrons. The chemical potentials in the 
{\em regular} region are $\lambda_p^{(1)}=4.35$MeV for the proton and 
$\lambda_n^{(1)}=3.66$MeV for the neutron. These values are almost constant within the region.

\subsection{Characteristic features of GCM solutions}

In Fig.6 we show behavior of the weight functions expressed by eq.(17) for several of the 
GCM solutions which are obtained from the constrained Hill-Wheeler equation (12). 
In the calculation, the number of states obtained by eq.(4) which are taken into account 
in solving eq.(12) are ``32''. That point locates within the {\em regular} region and is 
indicated in Fig.3b. From Fig.6 we confirm that the GCM solutions have symmetric or 
anti-symmetric properties with respect to the north and the south latitudes. This means 
that the broken symmetry of signature in TAR wave function has been recovered 
through the wobbling motion.  All the GCM solutions split into two groups, one is a 
symmetric group and the other is an anti-symmetric group. In the present case, the first, 
fourth, fifth, sixth and eighth solutions belong to the symmetric group and the second, third 
and seventh solutions belong to the anti-symmetric group. 

\begin{figure}
\psfig{figure=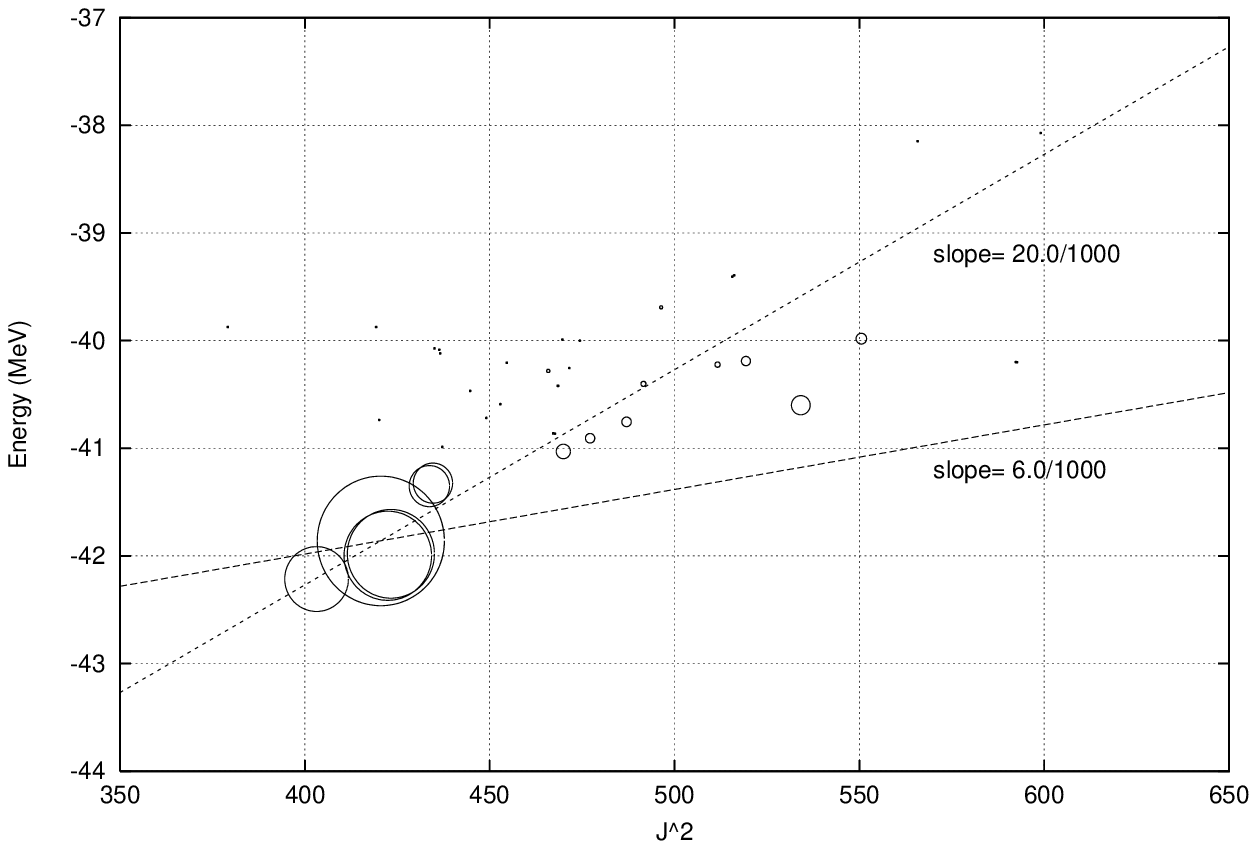,width=8.0cm}
\psfig{figure=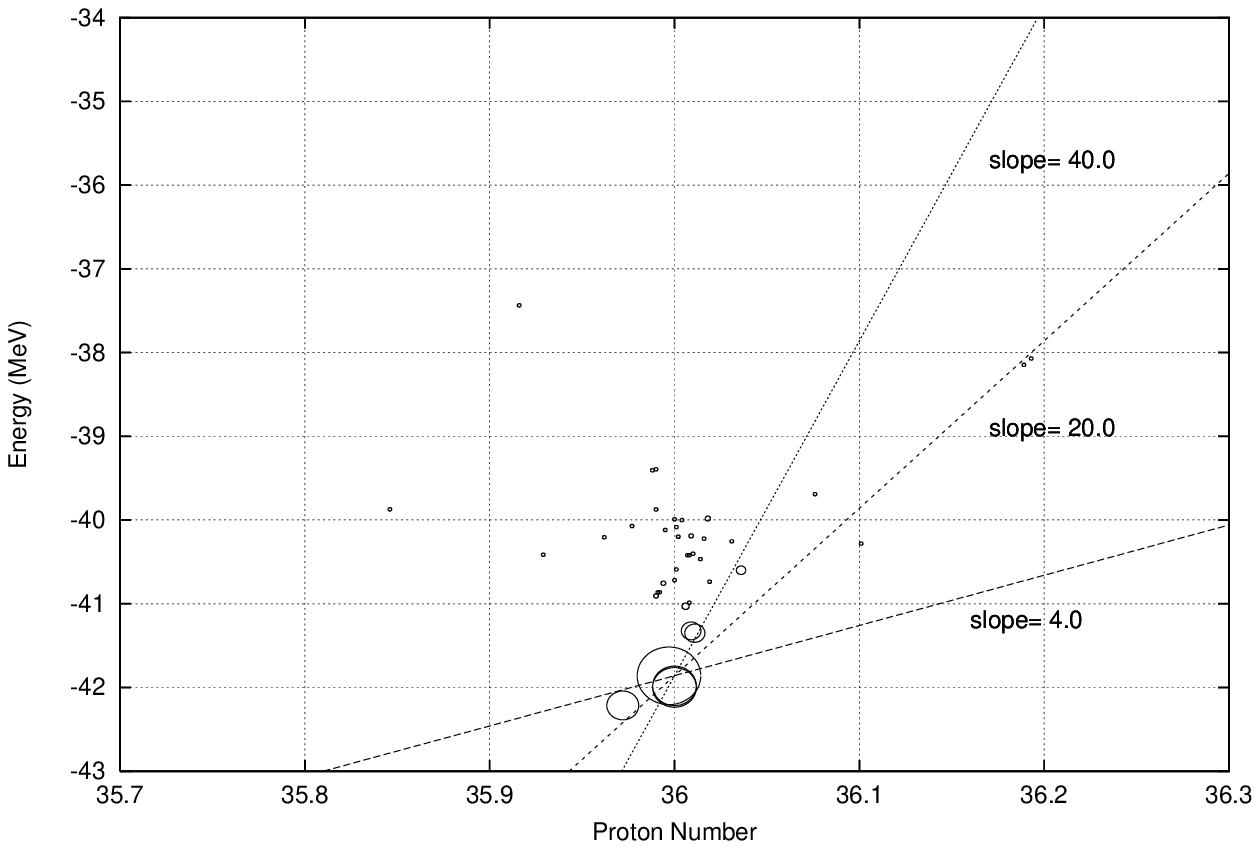,width=8.0cm}
\psfig{figure=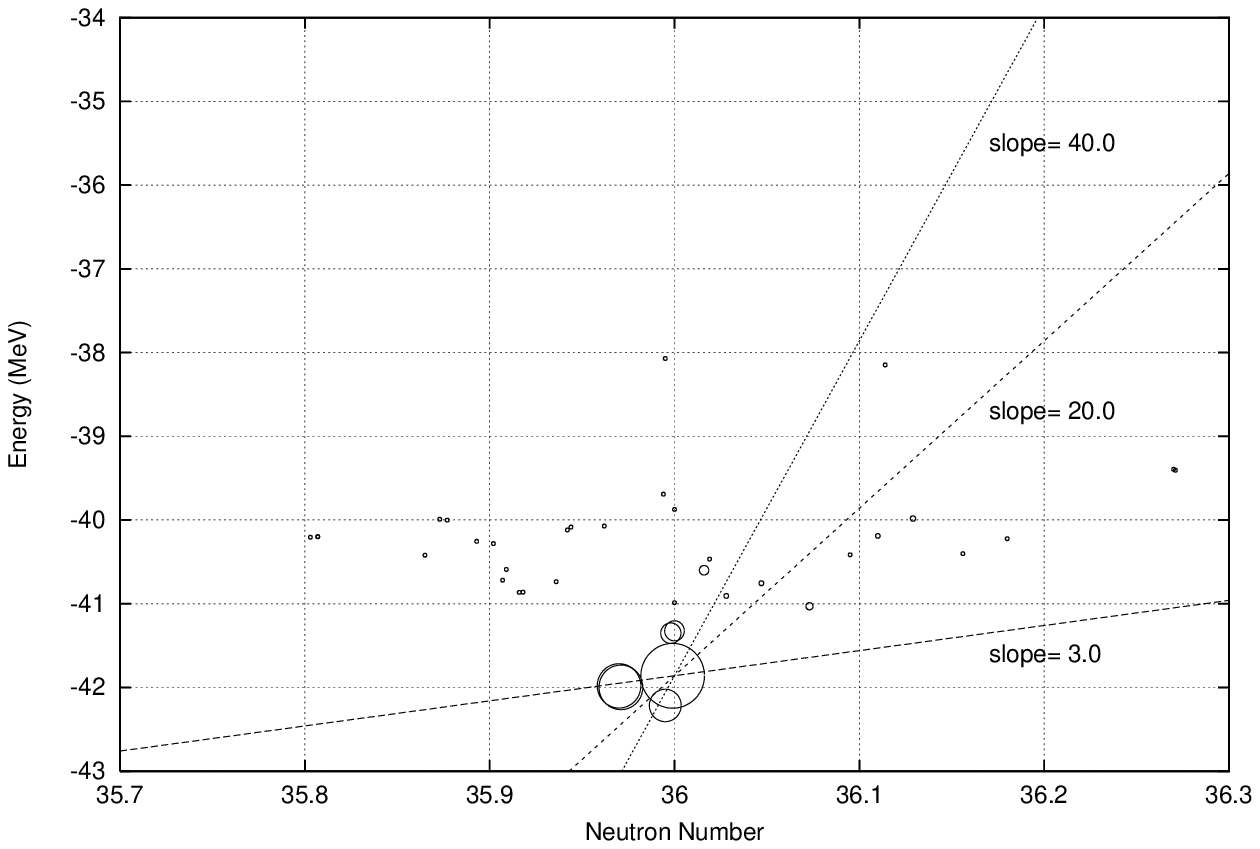,width=8.0cm}
\caption{The probability distribution of the GCM solutions. In the plane the radius
of each circle indicates the size of eigenvalue of the equation of norm overlap kernel.
The size express the probability of each solution. From the slope of this plot we can 
reduce the chemical potentials $\mu^{(1)}$, $\lambda_p^{(1)}$ and $\lambda_n^{(1)}$ which 
appear in eq.(12).}
\end{figure}


\begin{figure}
\psfig{figure=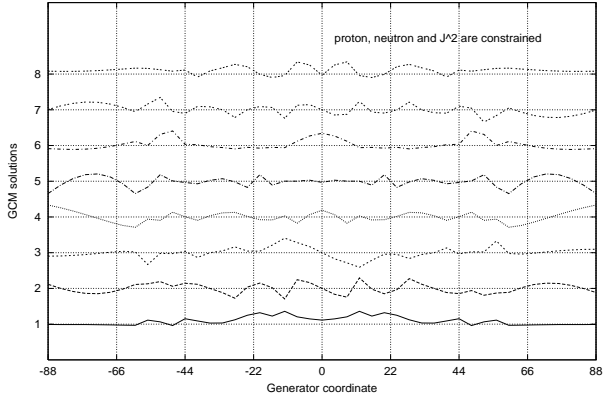,width=8cm}
\caption{The weight functions for the several lowest 
solutions of constrained Hill-Wheeler equation (eq.(12)) at the point indicated in Fig.3b. 
We can see that the GCM solutions have symmetric or anti-symmetric properties with respect 
to the north and the south latitudes. This means that the broken symmetry of signature in 
TAR wave function has been recovered through the wobbling motion.}
\end{figure}

The lowest solution expresses that this nucleus in the yrast $J=18\hbar$ state is a 
tilted-rotor at least for the present set of parameters. The tilt-angle is in the range of 
$12^{\circ} \sim 22^{\circ}$. The most probable angle is $\psi=12^{\circ}$. 
In Fig.2a we show the TAR potential curve which has a minimum point at $\psi=24^{\circ}$.
There is no guarantee, however, that a nuclear rotational motion will take place about
an axis indicated by this angle. In the present approach, only the GCM solutions 
based on the potential curve can properly describe the characteristic features of nuclear 
collective motion, such as a tilted axis rotation. This point was not clearly discussed
in our previous papers\cite{HO94,HOO95,HO96}.
Our lowest solution is symmetric with respect to $\psi=0^{\circ}$ and is thus 
characterized as {\em gerade} state. This solution also exhibits a vestige 
of a PAR character since it shows a small amplitude at $\psi=0^{\circ}$. 
As already discussed in section 3.1, the PAR state at $J=18\hbar$ in this nucleus 
is characterized by the almost vanishing neutron gap parameter and the still 
surviving proton gap parameter. Therefore we expect that this small PAR component of 
the state shares a neutron aligned $s$-band character. We call such state as an 
$s_n$-band state. These PAR characteristic features become remarkable in other 
solutions obtained at higher excitation energies as discussed below.  

The second and the third solutions also show the feature of the tilt states. 
The most probable tilt-angle for both states is also $\psi=12^{\circ}$. 
The second one, however, shows the tilt characteristic feature also at 
$\psi=8^{\circ}$. These two solutions have no amplitudes at the origin and show 
anti-symmetric character with respect to $\psi=0^{\circ}$. Thus they are grouped 
into the {\em ungerade} states. 

The fourth and sixth states are the {\em gerade} states and indicate a nature 
of the PAR state because they show certain amplitudes at $\psi=0^{\circ}$. 
This result is very interesting because we have solved the GCM equations based on 
the TAR potential curve which has the minimum points located at a distance from 
the principal axis. Since the state with an angular momentum $J=18\hbar$ is located  
in the neighborhood of the band crossing point in $^{182}$Os, we can naturally 
expect states with an $s$-band character in the GCM solutions. If above 
symmetric solutions are the member of an $s_n$-band, understanding of the backbending 
phenomena in this nucleus becomes quite interesting. 
P.M.Walker et al.\cite{W93} 
suggested that the backbending is expected to occur due to crossing 
of a $g$-band with a $t$-band and an $s$-band is not expected to come into play 
in their study. However, if such a PAR-like solution will come into the lowest state,
that is an yrast, in the higher angular momentum region, the situation warrants a
somewhat different scenario. The possibility of $s$-band
crossing after the $t$-band crossing emerges as one of the backbending mechanisms in
this nucleus. Therefore, we have found a sign of multi-band crossing in this nucleus. 

The fifth solution also turns out to be a tilt state whose tilt angle is 
$\psi=20^{\circ}$. Since the amplitude of the weight function between these two
tilt points is almost negligible, this state may be considered as almost a 
pure tilted-rotor. 

By looking at the higher angle region in Fig.6, we can find that 
all the lowest seven solutions show certain amplitude in the range of 
$\psi=44^{\circ}\sim66^{\circ}$. This result suggests an existence of a high-$K$ 
states ($K=12\sim16$). 
Further interesting features in the results are that the fourth and the fifth 
solutions show considerable amplitudes near the region of the north pole. This fact 
indicates that these solutions also exhibit a character of the deformation 
aligned state. The total angular momentum $J=18\hbar$ consists of $J_{n^+}=
8\hbar$ and $J_{p^-}=10\hbar$.  In experiments there are some discussions
about the existence of $K=8^{+}$ band in osmium\cite{CHO88,KLS95} and 
tungsten isotopes\cite{W93} in connection with an existence of the $t$-band. 
The interpretation of these deformation aligned states requires more precise 
analyses theoretically as well as experimentally.

\section{Summary and conclusions}
We have carried out the angular momentum and the particle number constrained
GCM calculation based on the three-dimensional cranked Hartree-Fock-Bogoliubov
approach. Compared to our previous work we have refined the force 
parameters carefully so as to reproduce the characteristic features of the
yrast states of the mass region. With our new set of force parameters the neutron 
gap parameter $\Delta_n$ almost vanishes whereas the proton gap parameter 
$\Delta_p$ still survives with the size of the ground state one at $J=18\hbar$. 
In our calculation the band crossing occurs at $J=20\hbar$. 
After choosing the force parameter set we have performed the 3D-CHFB calculation 
along the prime meridian on the globe of $J=18\hbar$ sphere. 
We know from our previous study that a calculation along 
the prime meridian always gives lower energy when compared with the calculation 
performed on the whole surface of the globe. 
Like the previous calculation, we have confirmed the 
existence of a stable TAR potential curve this time also. The TAR curve 
appears at $\psi=24^\circ$. Furthermore, here we have found a negative 
potential curve near the north pole. We have performed the GCM calculation based on 
this new potential curve and discussed the band structure properties of $^{182}$Os. 
We summarize the present work by stressing following three points. 

First point is that by including the constrained terms in the Hill-Wheeler equation 
the stability of the solutions is much increased and we can recognize in the GCM 
solutions the appearance of the wide range of plateaus even in such a high angular 
momentum region as $J=18\hbar$. This was not the situation in our previous paper\cite{HOO95}.
The reason is that we can suppress a drift of the mean value of the square of 
angular momentum value by the constrained term in the Hill-Wheeler equation and 
prevent a contamination by the lower angular momentum components present in the cranking
wave function. 

Second point is that from the symmetry property of the lowest GCM solution we can expect that 
this nucleus should be a tilted-rotor at $J=18\hbar$ yrast state. The tilt angle is about 
$\psi = 12^{\circ}$. As mentioned in section 3.3 the TAR potential curve will not be a 
necessary condition for an occurrence of the tilted axis rotation. In our approach only 
the GCM solutions properly predict the characteristic features of exotic collective motions.   
In the fourth and the sixth solutions we have found an $s$-band character in the PAR region. 
We have found a sign of multi-band mixing in this nucleus. If the $s_n$-band states, such as 
the fourth or the sixth solutions, come into energetically lower positions in the higher 
angular momentum region, we can say that the band mixing\cite{W93-2} will become a key 
concept in understanding the band structure of the nucleus.

The last point is that we have found the new 3D-CHFB solutions along the prime meridian 
which have been obtained by the tilt-back calculation at $J=18\hbar$ state. 
From the characteristic features of the solution, we can say that a new type of seesaw 
vibrational mode of the proton and the neutron pairing occurs in the backbending region 
in $^{182}$Os. This mode accompanies the large $\gamma$-oscillations. Thus a wobbling 
motion is expected to mediate a band crossing between the $g$-band and the $s$-band in 
this nucleus. The GCM calculation which includes these new solutions to explore the band 
structure of $A \sim 180$ mass region will be the next attractive subject.

\vspace{0.8cm}
\noindent
{\large\bf Acknowledgments}\\
The numerical calculations presented in the paper were carried out by
the Vector Parallel Processor, Fujitsu VPP500/28 at RIKEN.
The authors would like to express their thanks to Dr.\ S.Yamaji and Dr.\ Y.Yano 
for kindly letting them use the facilities in the Nishina memorial building 
at RIKEN. One of us (AA) would like to thank the Japan Society for the Promotion
of Science for financial support.
%

%
%
%

\begin{thebibliography}{10}
\bibitem{HW53}    D.L. Hill and J.A. Wheeler, Phys. Rev. {\bf 89} (1953) 1102.
\bibitem{CHO88}   P. Chowdhury et al., Nucl. Phys. {\bf A485} (1988) 136.
\bibitem{HO94}    T.Horibata and N.Onishi, Phys. Lett. {\bf B325} (1994) 283.
\bibitem{HOO95}   T.Horibata, M.Oi and N.Onishi, Phys. Lett. {\bf B355} (1995) 433.
\bibitem{ACCREP1} T.Horibata, M.Oi and N.Onishi, RIKEN Accel. Prog. Rep. {\bf 30} (1997) 20.
\bibitem{OOTH98}  M.Oi, N.Onishi, N.Tajima and T.Horibata Phys. Lett. {\bf B418} (1997) 1.
\bibitem{Bo90}    P. Bonche et al., Nucl. Phys. {\bf A510} (1990) 466.
\bibitem{HOO-I}   T.Horibata, M.Oi and N.Onishi, J. Phys. {\em to be published}.
\bibitem{HO96}    T.Horibata and N.Onishi, Nucl. Phys. {\bf A596} (1996) 251.
\bibitem{KO81}    A.K. Kerman and N. Onishi, Nucl. Phys. {\bf A361} (1981) 179, 
                     N. Onishi, Nucl. Phys. {\bf A456} (1986) 279.
\bibitem{HHR82}   K. Hara, A. Hayashi and P. Ring, Nucl. Phys. {\bf A385} (1982) 14,
                     E.W\"ust, A.Ansari and U.Mosel, Nucl. Phys. {\bf A435} (1985) 477.
\bibitem{RS80}    P.Ring and P.Schuck, {\em The Nuclear Many-Body Problem}, 
                     Springer-Verlag, Heidelberg, 1980.
\bibitem{GL67}    C. Gustafson, I.L. Lamm, B. Nilsson and S.G. Nilsson, Arkiv. Fysik. 
                     {\bf 36} (1967) 613.
\bibitem{HO97}    T. Horibata and N. Onishi, RIKEN Accel. Prog. Rep. {\bf 31} (1997) 22.
\bibitem{W93}     P.M.Walker et al., Phys. Lett. {\bf B309} (1993) 17.
\bibitem{KLS95}   T.Kutsarova et al., Nucl. Phys. {\bf A587} (1995) 111.
\bibitem{W93-2}   P.M.Walker, Proc. Future of Nuclear Spectroscopy, Crete (1993).
\end{thebibliography}
\end{document}